\documentclass{raa_rb}           
\usepackage{graphicx}            
\usepackage{natbib}
\usepackage{amssymb,amsmath}
\bibpunct{(}{)}{;}{a}{}{,}

\usepackage{subcaption}
\usepackage{multirow}
\usepackage{enumerate}
\usepackage{makecell}
\usepackage[pagebackref=true]{hyperref}

\begin{document}

  \title{Long-term evolution of solar activity and prediction of the following solar cycles}
  
   \volnopage{Vol.0 (20xx) No.0, 000--000}      
   \setcounter{page}{1}          
   
   \author{Pei-Xin Luo 
      \inst{1,2}
   \and Bao-Lin Tan
      \inst{1,2}
   }
   
     \institute{National Astronomical Observatories of Chinese Academy of Sciences,
                Beijing 100101, China; {\it luopeixin23@mails.ucas.ac.cn}\\
        \and
             School of Astronomy and Space Science, University of Chinese Academy of Sciences, Beijing 100049, China \\ 
\vs\no
   {\small Received 20xx month day; accepted 20xx month day}}

\abstract{Solar activities have a great impact on modern high-tech systems, such as human aerospace, satellite communication and navigation, deep space exploration, and related scientific research. Therefore, studying the long - term evolution trend of solar activity and accurately predicting the future solar cycles is highly anticipated. Based on wavelet transform and empirical function fitting of the longest recorded data of the annual average relative sunspot number (ASN) series of 323 years to date, this work decisively verified the existence of the solar century cycles and confirmed that its length is about 104.0 years, and the magnitude has a slightly increasing trend on the time scale of several hundreds of years. Based on this long-term evolutionary trend, we predicted solar cycle 25 and 26 by using phase similar prediction methods. As for the solar cycle 25, its maximum ASN will be about $146.7\pm 33.40$, obviously stronger than solar cycle 24. The peak year will occur approximately in 2024, and the period is about $11\pm 1$ years. As for the solar cycle 26, it will start around 2030, reach the maximum between 2035 and 2036, with maximum ASN of about $133.0\pm 3.200$, and the period is about 10 years.  
}

   \authorrunning{P.-X. Luo\& B.-L. T }            
  \titlerunning{Long-term evolution and prediction of solar cycles}  

  \maketitle 
  
%
\section{Introduction}           
\label{sect:1}

The Sun, as the unique star in the solar system that is closest to the Earth, plays a crucial role in the formation and evolution of the Earth and its near - Earth space environment. At the same time, solar activities may produce a great damage to the modern high - tech human systems, such as aerospace, satellite communication and navigation, large power grids, deep - space exploration, national safety and the related scientific research. Studying the long - term trends in solar activity will help us answer the big questions such as how the Earth's future environment will change and where humans will go. At least, it can also help us predict the intensity and development trend of future solar activity.

As we know, the solar activity has cyclical characteristics, reflected in various indicators, such as sunspot numbers, sunspot areas and 10.7\,cm - wavelength solar radio flux, etc. One of the most obvious periodicities is the solar cycle with an average period of about 11 years, also known as the Schwabe cycle, first reported by \citet{schwabe1844}.

Over the past 200 years since the discovery of the Schwabe solar cycle, some evidence with other periodic characteristics has also been discovered. Such as the grand minimum, the G - O rules that show the difference between even and odd cycles, and the century cycle \citep{hathaway2015}. At present, some of these features are well understood, but others are not yet conclusive. For example, does the century cycle really exist? How long is its period? Is it a change? And how to change? 

The solar century cycle refers to a periodicity of approximately around one century, first proposed by \citet{gleissberg1939}. He smoothed the amplitude of sunspots numbers occurring between 1750 and 1928 and found a long period of 7 to 8 solar cycles, which is close to a century and also known as the secular cycle or Gleissberg cycle. In 1994, \citet{rozelot1994} used Fourier analysis and proposed that the period of the century cycle is 97.2 years. However, Garcia and Mouradian's analysis \citeyearpar{garcia1998} four years later suggested that the period of the century cycle is 78 or 81 years. Other researchers \citep{peristykh2003} obtained the evidence that the period should be 88 years by using the analysis of cosmogenic isotopes.  \citet{ogurtsov2002} more explicitly described the century cycle as a mixture of biperiodic structures with a 50 - 80 years period and a 90 - 140 years period. \citet{le2003} proposed that the period should be about 101 years. Recently, some researchers \citep{ma2009} have extended the length of the century cycle to between 60 and 150 years. \citet{tan2011} analyzed ASN during 1700 AD – 2009 AD and proposed that there should be 3 kinds of solar cycles: the well-known 11 - year Schwabe cycle, the 103 - year century cycle and the 51.5 - year cycle. The periodicity of the century cycle also appears in solar proton events and possibly linked to the grand minimum \citep{mccracken2001}.

The causes of the century cycle were analyzed. In 1999, \citet{pipin1999} proposed a numerical model of the century cycle based on the dynamo mechanism describing the generating processes of solar magnetic fields, and estimated that the century cycle is the result of the re-establishment of differential rotation after the magnetic feedback of angular momentum transport. Some researchers take a different view of the century cycle. In 1999, \citet{hathaway1999} used the century cycle to predict future cycles and found that the best-fitting value of the century cycle was constantly changing rather than stable. 

\citet{schwabe1843} made the first solar cycle prediction \citep{attia2020}. After almost two hundred years of development, \citet{petrovay2010} divided solar cycle prediction methods into three categories: methods based on physical models, experience-based precursor and extrapolation methods. 

The model-based method mainly refers to the dynamo model of the solar magnetic field. Precursor method relies on previous solar activity indicators or magnetic fields at earlier times to predict the maximum amplitude of the next solar cycle \citep{petrovay2020}. It is mainly divided into two categories: geomagnetic index prediction and polar magnetic field prediction. The prediction of the polar magnetic field as a precursor was first proposed by \citet{schatten1978}. \citet{nandy2021}believes that the precursor method is relatively superior to other methods in terms of accuracy, especially the approach that uses the polar field near the solay cycle minimum as the precursor factor. The use of the geomagnetic index as a precursor dates back to 1966. In that year, Ohl found that the value of the aa index indicating geomagnetic activity near the time of the solar cycle minimum correlated well with the amplitude of the next cycle \citep{hathaway2015}. According to the definition of the precursor method, not only the geomagnetic index and the magnitude of the polar magnetic field can be utilized as precursors, but also the sunspot numbers in the declining phase of the cycle \citep{brajsa2022, nagovitsyn2023}, the number of spotless days in the declining phase of the cycle \citep{burud2021}, etc. 

The extrapolation methods based on statistical analysis of various indicators of solar activity to obtain its mathematical law, then applied this law to reasonably extrapolate and predict the following solar cycles. Recently, this method has been used to predict the solar cycle 25 \citep{kakad2021}. There are many metrics that extrapolation method can be utilized, but the most widely used is the sunspot number sequence. Sunspots have been observed for more than three centuries and should be the longest lasting time series in the world \citep{bhowmik2018}, which is one of the main reasons why the sunspot sequence was chosen as a predictor in this work. Additionally, similarity theory also can be used to predict the forthcoming cycles which assumed that two solar cycles with similar solar activity indicators should also have similar features \citep{du2023}. 

In addition to the above classification, cross-disciplinary approaches such as machine learning and neural networks for solar cycle prediction are also very popular research directions.\citet{ramos2023} summarizes the results on using machine learning to make predictions of solar cycle 25.

The methodology of this work is based on extrapolation method and similarity theory. Firstly, we confirm the existence and long - term evolution trend of the solar century cycle, obtain an empirical century cycle function by using fitting method. Then we proposed that a 11 - year solar cycle at the similar phase in the century cycle should have similar characteristics (such as length, magnitude, etc.). And then we make predictions for the current solar cycle 25 and the upcoming solar cycle 26. Section 2 describes in detail the data and method used in this work. The main results of the long-term evolution of solar activity and the prediction of the forthcoming solar cycles, as comparisons with the other results are presented in Section 3, and finally, the conclusions of this work are summarized in Section 4.

\section{Data and Methodology}
\label{sect:2}
  
The data used in this work are ASN and the 13 - month smoothed monthly total sunspot number which are compiled and released by the World Data Center (WDC) - Sunspot Index and Long - Term Solar Observations (SILSO), Royal Observatory of Belgium, Brussels. It can be downloaded from the website: \url{www.sidc.be/SILSO/datafiles}, which includes 323 years of ASN data recorded from 1700 AD to present. This is the longest continuous solar activity time series available to date, and is one of the most important basic data for studies of the long - term evolution of solar activity.
  
In order to better reflect the long-term trend of the cycle maximum of ASN of solar cycles, we first assumes a sinusoidal empirical function, which can be expressed as: 
\begin{equation}\label{eq1}    
R=A+B\sin(\frac{2\pi x}{C}+D\pi)+Ex           
\end{equation}

Here $x$ represents the years after 1700 AD, and $R$ represents the values of ASN of corresponding to that year. $A$, $B$, $C$, $D$, and $E$ are five unknown parameters that need to be determined by us. The assumed empirical function (Eq.~ \eqref{eq1}) is a nonlinear fit function, and the parameters are determined in this paper using mathematical analysis software. Here, the constant $A$ represents a level of the background, $B$ is the magnitude of the century cycle, $C$ is the period, $D$ represents the initial phase, and $E$ reflects a long - term trend of the century cycles. Obviously, the century cycle is quasi - periodic. For convenience, we firstly determined the parameter $C$ and $D$ before fitting the function. The wavelet analysis method is used to analyze the ASN sequence and to determine whether the century cycle ends or not by combining the actual data with the consideration of the quasi - periodicity of solar cycles, and finally to determine the value of parameter $C$ and $D$.

\citet{tan2011} found that $E=0.06$, which indicates a slow increasing trend in solar activity on the century timescale. \citet{javaraiah2017a} also found that the maximum of solar cycles has a linear tendency to increase, and it has also been argued that there is no tendency for the maximum of the solar cycle to increase since the Maunder Minimum \citep{svalgaard2011}. Therefore, the empirical function assumed in this work also contains the item of linear variation, and its coefficient is denoted by $E$. The function is fitted to determine whether there is a trend for the maximum of the solar cycle to increase ($E>0$) or decrease ($E<0$).

This work assumes that a Schwabe cycle locating at the same or similar phases in the century cycle will have the same or similar cycle characteristics, such as cycle maximum, rise time, and cycle length. Then, we may adopt the phase similarity method to predict the basic features of solar cycle 25 and 26. The specific steps are as follows: (1) determine the empirical function corresponding to the century cycle, (2) extend the fitting function forward to determine the phases of solar cycle 25 and 26 in the century cycle, (3) extend the fitting function backward to determine the previous cycles that have the same or similar phase with them, (4) compare solar cycle 25 and 26 with their corresponding similar cycles and determine their predicting characteristic parameters.

\begin{figure}[ht]
  \centering
  \begin{subfigure}[c]{\textwidth}
    \includegraphics[width=\textwidth]{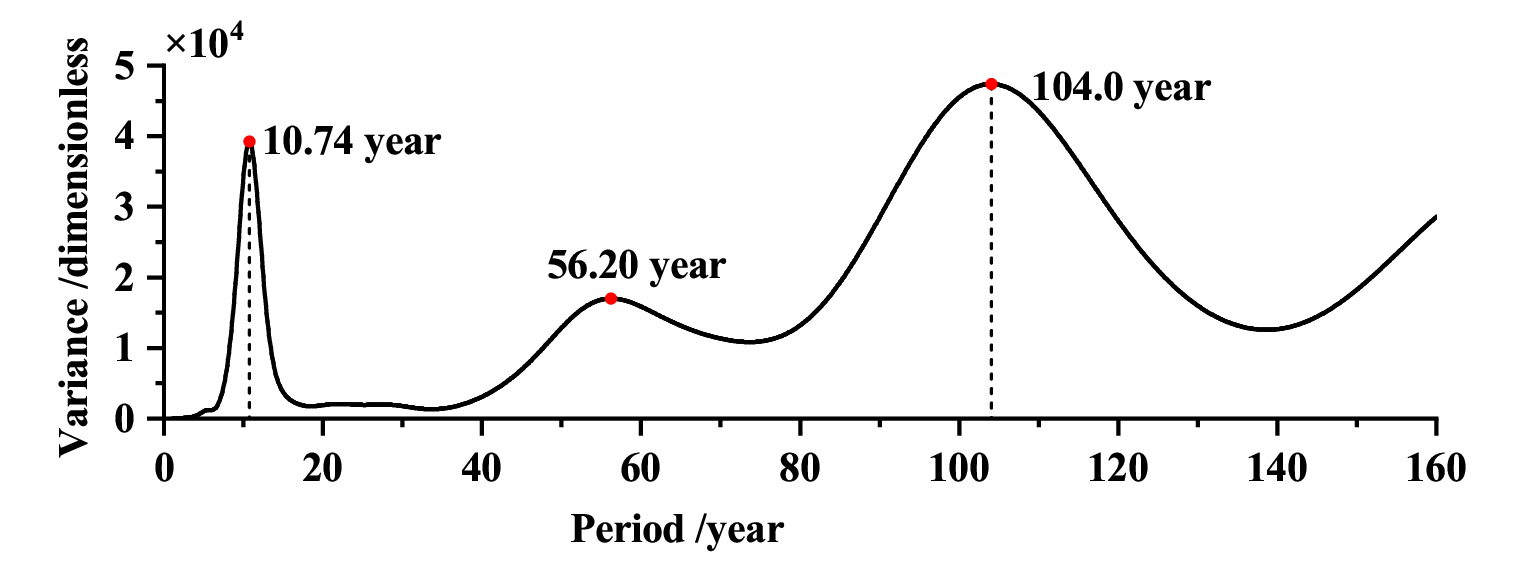}
    \caption{}
    \label{fig1a}
  \end{subfigure}
  \begin{subfigure}[c]{\textwidth}
    \includegraphics[width=\textwidth]{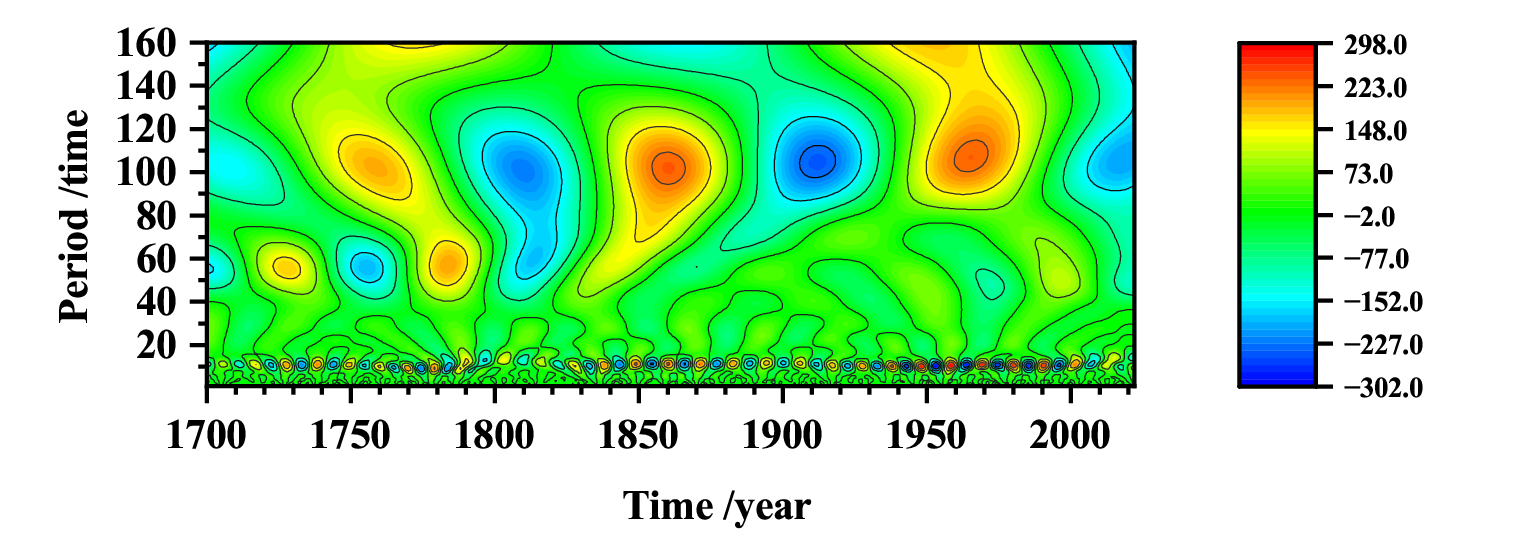}
    \caption{}
    \label{fig1b}
  \end{subfigure}
  \caption{Figure 1a shows the variance plot of the wavelet coefficients after the wavelet transform of the ASN sequence, Figure 1b shows the contour plot of the real part of the wavelet coefficients after the wavelet transform of the ASN sequence.}
  \label{fig1}
\end{figure}
   
\section{Results and Discussion}
\label{sect:3}

\subsection{Result}
\subsubsection{Determination of the Century Cycle}

Here, we choose the wavelet transform method to analyze ASN time series. To ensure the analysis accuracy, we subtract the mean from the original ASN time series and then did a symmetric extension of the data on both sides to reduce boundary effects.

Figure \ref{fig1a} shows that the tendency of the variance to change with the period. Here, we can find that variance at about 104.0 - year period is the maximum in all periods, which means that the 104.0 - year period is the most significant periodic component of the solar activity. Furthermore, we also find that in terms of peak sharpness, it is obvious that the peak of the solar cycle is sharper, while the peak of the 104.0 - year period is wider, which indicates that the quasi - periodic characteristic of the 104.0 - year cycle is stronger. The timescale of the solar cycle obtained by wavelet analysis is slightly less than the well-known 11 years. Due to the quasi-periodic characteristic of the solar cycle, we believe that the value is within the reasonable range of the timescale of the solar cycle. If we assume that the duration values of the 29 complete solar cycles observed since 1700 AD conform to a normal distribution and perform a Gaussian function fitting, the mean value is around 10.78, which is close to the value we obtained. The third obvious crest is the 56.20 - year period, numerically close to half of the century cycle. In order to more intuitively describe the periodicity of ASN time series, Figure \ref{fig1b} shows the contour plot of the real part of the wavelet coefficients of ASN time series. The contour plot of the real part of the wavelet coefficients can reflect the periodical changes of ASN time series in different period and its distribution in the time domain, and then can judge the future trend of ASN in different period. As can be seen in Figure \ref{fig1b}, there are multiple cycle components in ASN time series during the evolution from 1700 AD to 2022 AD. The cycle components of 10.74 and 104.0 years are consistently present, whereas the 56.20 - year period is only partially significant on the time axis. This hints at the possibility that the 56.20 - year cycle is not a stable periodic feature of solar activity.

\begin{figure}[ht]
   \centering
   \includegraphics[width=\textwidth, angle=0]{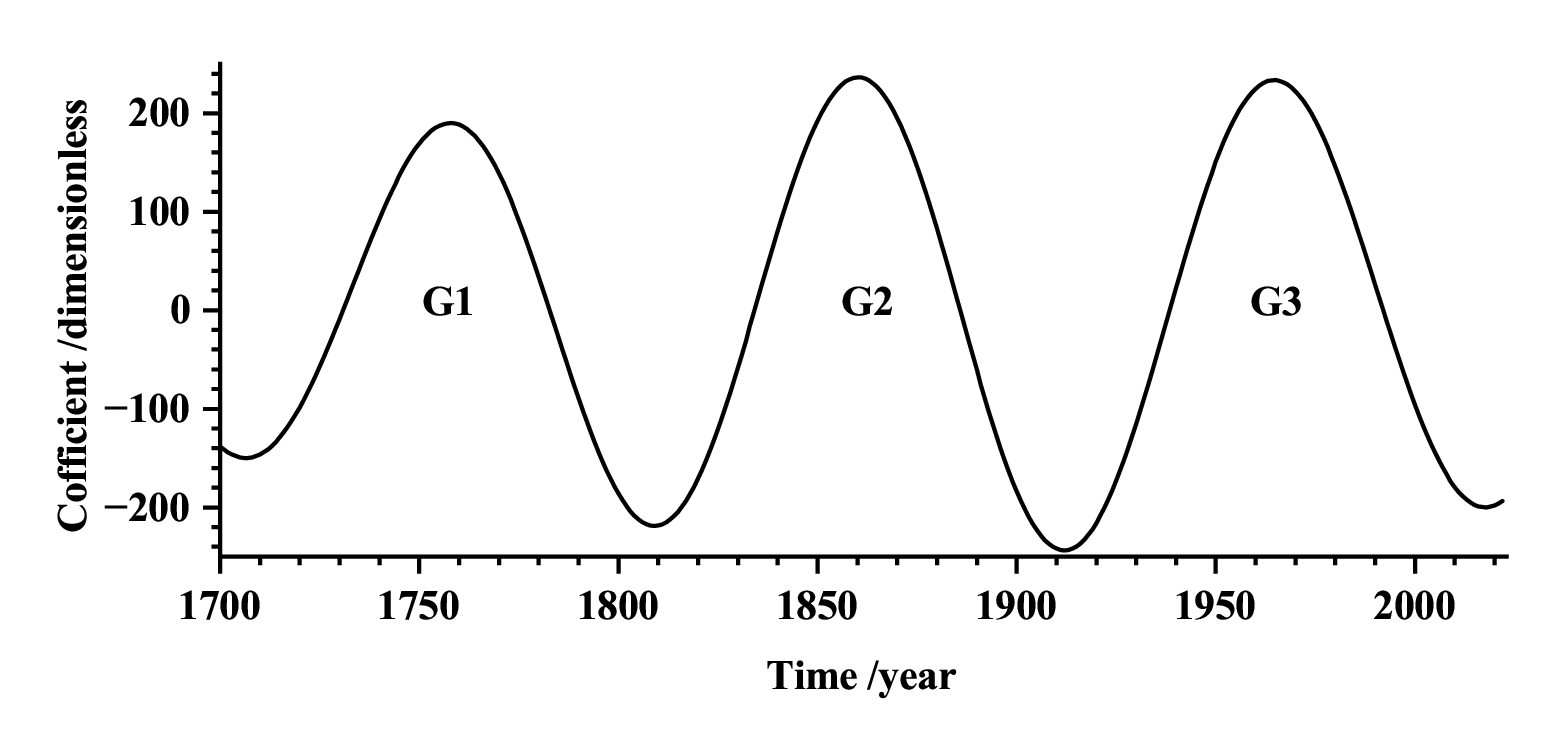}
   \caption{The variation of wavelet coefficients in the time domain at 104.0 - year period. }
   \label{fig2}
   \end{figure}

The wavelet coefficients at 104.0 - year period are plotted in Figure \ref{fig2}. We find that there are 3 full cycles. In addition, the absolute amplitude in the Figure \ref{fig2} increases gradually with time, which also indicates that there is a gradual trend of increasing solar activity since 1700 AD. 

Different works have come to different results on the timescale of the century cycle \citep{gleissberg1939,garcia1998,peristykh2003}. However, based on our analysis with the longest data series so far in this work, we believe that the more accurate timescale of the century cycle should be 104.0 years, which is close to the results of 101 years \citep{le2003}, 103 years \citep{tan2011}, and 100 years\citep{feynman2014}, and far from the results of 90 years \citep{le2017a}. 

After determining the timescale of the century cycle, we need more precise confirmation of the initial phase of the century cycle on ASN time series to fit the empirical trend equation. 

\begin{figure}[ht]
   \centering
   \includegraphics[width=\textwidth, angle=0]{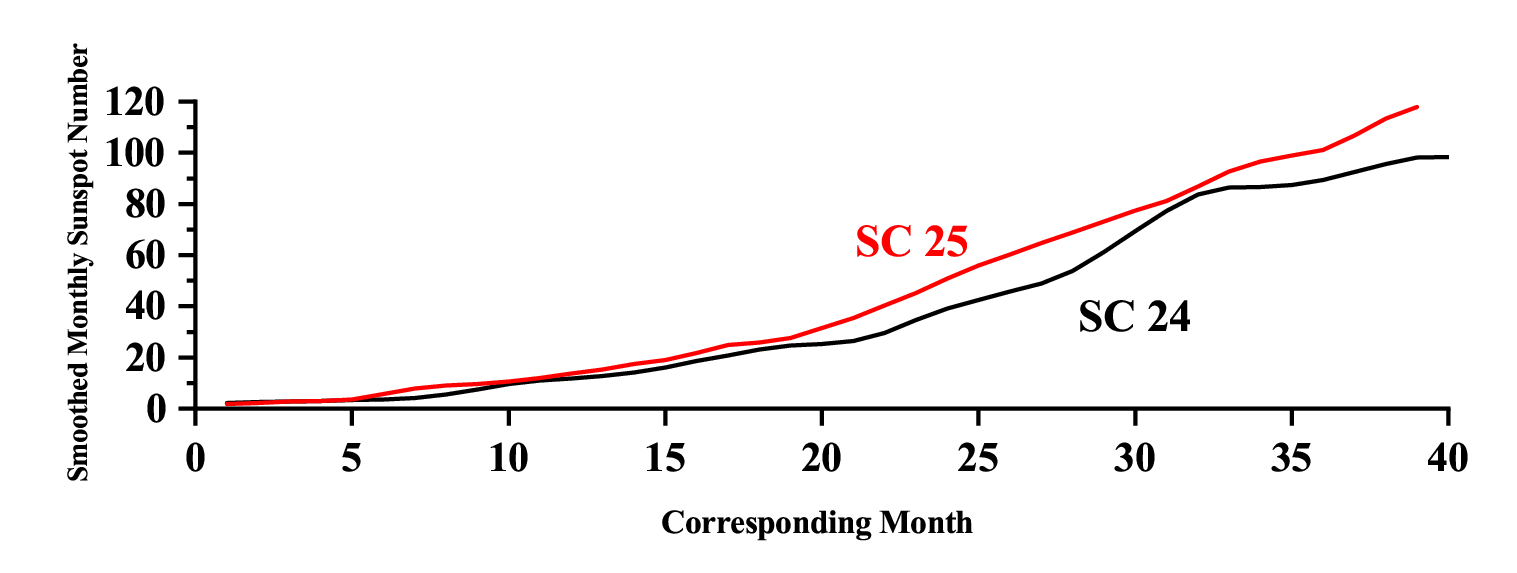}
   \caption{The red curve presents the smoothed monthly mean sunspot number during the rising phase of solar cycle 25 (from 2019 December to 2023 April), while the black curve represents the results during the rising phase of solar cycle 24 (from 2008 December to 2012 April). }
   \label{fig3}
   \end{figure}

In an earlier study of the century cycle, \citet{tan2011} suggested that solar cycle 24 is located in the valley between the 3rd and 4th century cycles. We compare the 13 - month smoothed monthly total sunspot numbers of the corresponding months in the rising phase of solar cycle 24 and 25 one by one. The start of the two solar cycles begins with the month in which the minimum of the 13 - months smoothed monthly total sunspot numbers occurs, 2008 December and 2019 December, respectively. The results are shown in Figure \ref{fig3}. We find that the 13 - month smoothed monthly total sunspot numbers for solar cycle 25, as of 2023 February, is overall greater than the corresponding sunspot numbers for solar cycle 24. In other words, solar cycle 25 is significantly stronger than solar cycle 24. 

Then, how to determine the end time of the last century cycle (G3) and the start time of the forthcoming century cycle (G4)? From our above analytical results, we know that the solar cycle 24 is the weakest cycle in the past half century, and the current solar cycle 25 is obviously stronger than solar cycle 24. This fact strongly indicates that solar cycle 24 is precisely located at the lowest point between G3 and G4. Therefore, we may conclude that the end of G3 and the beginning of G4 occurs in the middle of solar cycle 24, that is to say, around 2014 April. And solar cycle 25 is already in the rising phase of the forthcoming century cycle G4. Going back in time, we can see that 1702 AD is the starting year of the first full century cycle to appear in the ASN time series.

\subsubsection{Fitting of Hypothesized Empirical Function}

This paper focuses on the long - term trend of solar activity, known as the century cycle, and utilizes it to predict the parameters of the forthcoming solar cycles. Therefore, in order to prevent our study from being affected by outliers in ASN time series, we smoothed ASN time series to better emphasize the long-term trend: the maximum of each solar cycle is averaged with its minimum at the beginning and with its minimum at the end. The resulting averages are used as elements of the smoothed ASN time series, and the corresponding time are the closest years obtained by indexing real ASN.

\begin{figure}[ht]
   \centering
   \includegraphics[width=\textwidth, angle=0]{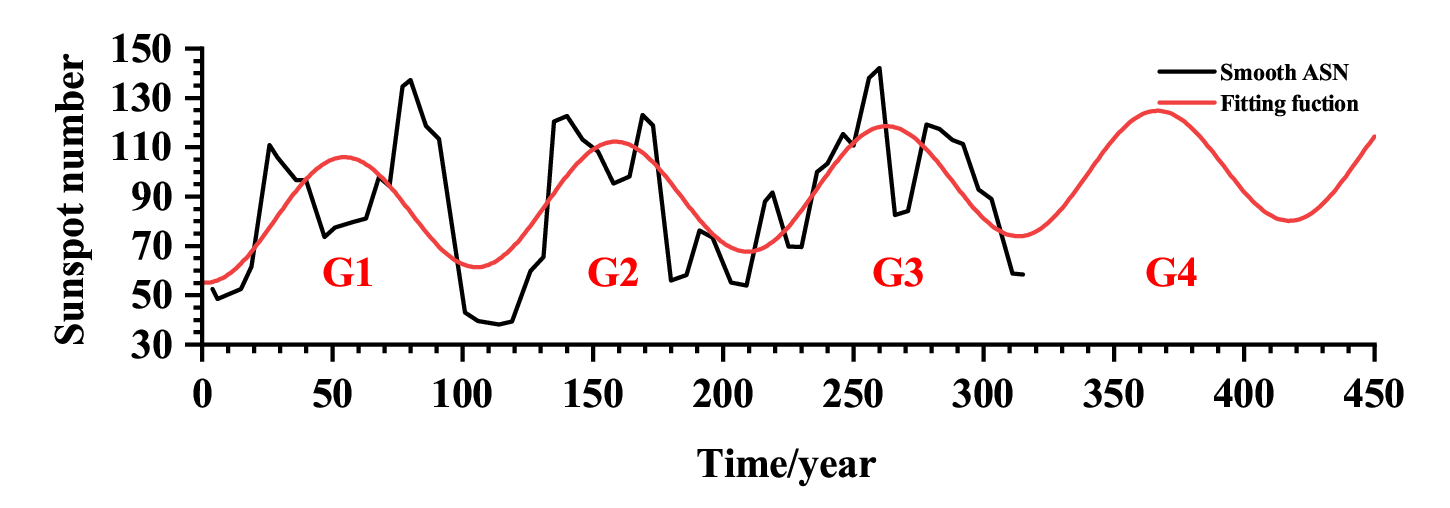}
   \caption{The black curve represents the smoothed ASN series over time, and the red curve is the fitted function of the century cycles. }
   \label{fig4}
   \end{figure}

The ASN sequence we used to be recorded from 1700 AD, but the existence of sunspots is much earlier than 1700 AD. So, in order to minimize the error, this paper takes 1700 AD as the starting point, with a total of 58 sets of data in terms of years, ending at the end of solar cycle 24. Its processed image is shown as the black curve in Figure \ref{fig4}. 

We assume that the empirical function in the form of Equation~ \eqref{eq1}, where the parameter $C$ is the length of the corresponding century cycle of 104.0 years, and parameter $D$, which can be calculated from the fact that 1702 is the starting year of the century cycle G1, has a value of -0.5385.

Then, by using the least squares fitting method, we obtained the values of parameters $A$, $B$, and $E$ in Equation~\eqref {eq1}, and the function can be expressed as:
\begin{equation}\label{eq2}
R=78.84+23.86sin(\frac{2\pi x}{104.0}-0.5385\pi)+0.06028x
\end{equation}

The coefficient of the linear term of the equation is 0.06028, which indicates that the intensity of solar activity tends to increase linearly on a time scale of centuries. From Figure \ref{fig4}, we can also clearly see the presence of the century cycle.

The fitted function is extended outward by 150 years, as shown by the red curve in Figure \ref{fig4}. Parameters of future solar cycles are inferred by the principle of extrapolation of similarity of the parameters of solar cycles in the same phase of the century cycle.

\subsubsection{Predictions for Solar Cycle 25 and 26}

The method in this paper is a phase - similar prediction based on the evolutionary trend of the century cycle, so before the prediction we have to make a collation of the parameters of all solar cycles so far.

The international numbering of solar cycles began in 1755, and there have been 24 full solar cycles, so far. In the period from 1700 to 1755, ASN data also went through five full solar cycles, which for the purposes of analysis we have named alphabetically from latest to earliest as A, B, C, D, and E \citep{tan2011}. 

In addition to the numbering, the characteristic parameters used to describe the solar cycle are as follows:
\begin{enumerate}[(a)]
\item The year $t_{0}$ of the start of the solar cycle (referred to as the start time);
\item Minimum of ASN at the start of the solar cycle, $R_{min}$;
\item The length of the rising phase of the solar cycle, $T_{up}$ (referred to as the rise time);
\item The year of the maximum of the solar cycle, $t_{max}$ (referred to as the peak time);
\item Maximum of ASN of the solar cycle, $R_{max}$;
\item The length of the decay phase of the solar cycle, $T_{de}$ (referred to as the fall time);
\item The period of the solar cycle, $L$ (referred to as the period);
\item Asymmetric parameter, $R_{at}$, the ratio between the rise time and the fall time, which represents whether the shape of the solar cycle is left - right symmetrical or not.
\end{enumerate}

The parametric characteristics of Schwabe cycles are listed in Table 1.

Table \ref{tab1} lists the parametric characteristics of 29 solar Schwabe cycles since 1700 AD. First of all, the period $L$ takes values between 9 and 14 years, and 82.76\% of the period $L$ of all recorded solar cycles is between 10 and 12 years. This confirms the quasi - cyclical nature of the solar cycles. Secondly, there's a big difference among the maxima of the different solar cycles. So far, the smallest observed maximum occurred during solar cycle 6 with maximum ASN of 76.30. The highest maximum occurred during solar cycle 19 with maximum ASN of 269.3. In addition, the following conditions are observed in the asymmetric parameter $R_{at}$: 75.86\% of $R_{at}$ of solar cycles are less than 1, 13.79\% of $R_{at}$ of solar cycles are greater than 1 and 10.34\% of $R_{at}$ of solar cycles are equal to 1. This indicates that most solar cycles are rapidly rising and slowly falling, belonging to the left-biased asymmetric type.

After clarifying the main properties of the solar cycles, we try to make a prediction of the forthcoming solar cycle 25 and 26.

\begin{figure}[ht]
   \centering
   \includegraphics[width=\textwidth, angle=0]{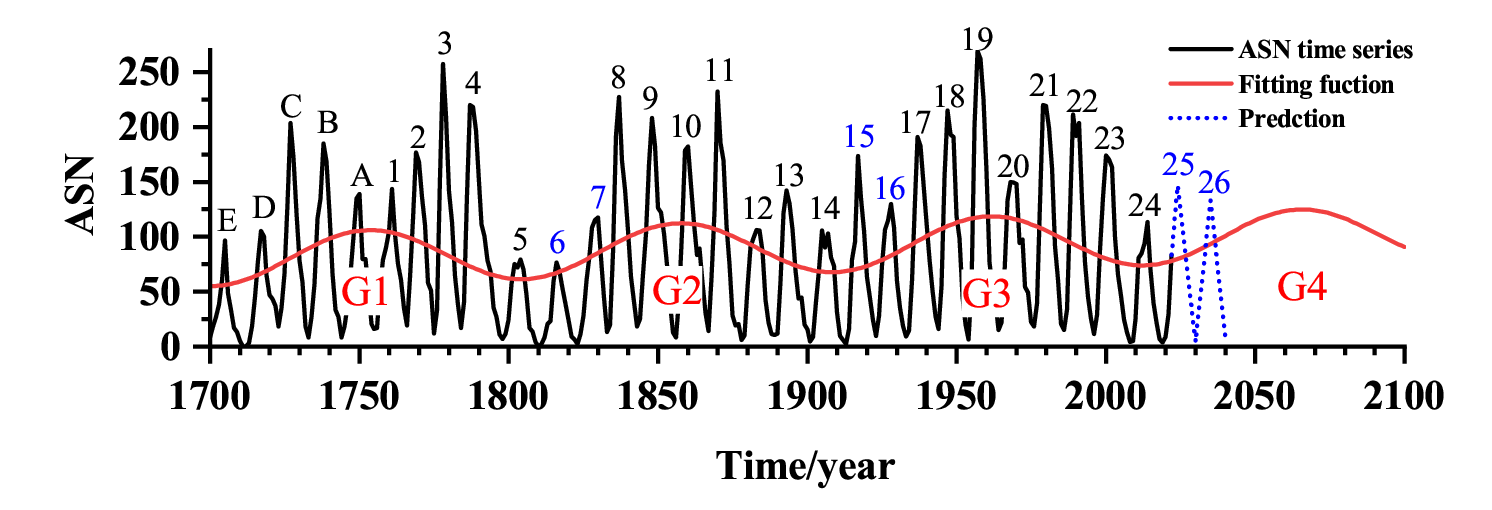}
   \caption{The black dotted curve represents the trend of the observed ASN with time, and A - E and 1 - 26 are numbers of the Schwabe solar cycles. The red curve represents the fitted curve of the century cycles, G1, G2, G3, G4 numbered the different century cycles, respectively. The blue dashed lines are the predicted profiles of the forthcoming solar cycle 25 and 26. }
   \label{fig5}
 \end{figure}

Firstly, let's determine the phase of the upcoming solar cycle 25 and 26 in the century cycle and other Schwabe cycles with similar phases. Figure \ref{fig5} plotted the temporal profiles of 29 Schwabe cycles numbered with 1 - 24 and A - E and overlaid the century cycles numbered G1 – G4 since 1700, respectively. It shows that the solar cycle 24 is located at the valley between the century cycles G3 and G4, while the solar cycle 25 and 26 are the first and second cycle in the ascending phase of the century cycle G4. Considering the solar cycles located in the early part of the century cycle G1 (A, B, C, D, and E), for which there is a large amount of missing data \citep{clette2015,clette2016}, we do not consider solar cycles located in the early part of the century cycle G1 when searching for the corresponding past solar cycles.

We find that the phase of solar cycle 25 is very similar to that of solar cycle 6 in the century cycle G2 and solar cycle 15 in the century cycle G3, while the phase of solar cycle 26 is very similar to that of solar cycle 7 in the century cycle G2 and solar cycle 16 in the century cycle G3. Naturally, we may apply the averaged parameters’ values of solar cycle 6 and 15 to predict the main characteristics of solar cycle 25, and apply the averaged parameters’ values of solar cycle 7 and 16 to predict the main characteristics of solar cycle 26. At the same time, we have to consider the effect of the last linear increasing term of the Equation~ \eqref{eq2}.

\begin{table}[ht]
\bc
\begin{minipage}[]{100mm}
\caption[]{Summary of Solar Cycle Parameters\label{tab1}}
\end{minipage}
\setlength{\tabcolsep}{1pt}
\small
 \begin{tabular*}{\textwidth}{@{\extracolsep{\fill}}c c c c c c c c c@{}}
  \hline\noalign{\smallskip}
Numbering& $t_{0}$ & $R_{min}$ & $T_{up}$ & $t_{max}$ & $R_{max}$ & $T_{de}$ & $L$ & $R_{at}$ \\
 &(year)& &(year)&(year)&&(year)&(year)&\\
  \hline\noalign{\smallskip}
E	&1700	&8.300	&5	&1705	&96.70	&6	&11	&0.8330 \\
D	&1711	&0.000	&6	&1717	&105.0	&6	&12	&1.000 \\
C	&1723	&18.30	&4	&1727	&203.3	&6	&10	&0.6670 \\
B	&1733	&8.300	&5	&1738	&185.0	&6	&11	&0.8330 \\
A	&1744	&8.300	&6	&1750	&139.0	&5	&11	&1.200 \\
1	&1755	&16.00	&6	&1761	&143.2	&5	&11	&1.200 \\
2	&1766	&19.00	&3	&1769	&176.8	&6	&9	&0.5000 \\
3	&1775	&11.70	&3	&1778	&257.3	&6	&9	&0.5000 \\
4	&1784	&17.00	&3	&1787	&220.0	&11	&14	&0.2720 \\
5	&1798	&6.800	&6	&1804	&79.20	&6	&12	&1.000 \\
6	&1810	&0.000	&6	&1816	&76.30	&7	&13	&0.8570 \\
7	&1823	&2.200	&7	&1830	&117.4	&3	&10	&2.333 \\
8	&1833	&13.40	&4	&1837	&227.3	&6	&10	&0.6670 \\
9	&1843	&18.10	&5	&1848	&208.3	&8	&13	&0.6250 \\ 
10  &1856	&8.200	&4	&1860	&182.2	&7	&11	&0.5710 \\
11	&1867	&13.90	&3	&1870	&232.0	&8	&11	&0.3750 \\
12	&1878	&5.700	&5	&1883	&106.1	&6	&11	&0.8330 \\
13	&1889	&10.40	&4	&1893	&142.0	&8	&12	&0.5000 \\
14	&1901	&4.600	&4	&1905	&105.5	&8	&12	&0.5000 \\
15	&1913	&2.400	&4	&1917	&173.6	&6	&10	&0.6670 \\
16	&1923	&9.700	&5	&1928	&129.7	&5	&10	&1.000 \\
17	&1933	&9.200	&4	&1937	&190.6	&7	&11	&0.5710 \\
18	&1944	&16.10	&3	&1947	&214.7	&7	&10	&0.4290 \\
19	&1954	&6.600	&3	&1957	&269.3	&7	&10	&0.4290 \\
20	&1964	&15.00	&4	&1968	&150.0	&8	&12	&0.5000 \\
21	&1976	&18.40	&3	&1979	&220.1	&7	&10	&0.4290 \\
22	&1986	&14.80	&3	&1989	&211.1	&7	&10	&0.4290 \\
23	&1996	&11.60	&4	&2000	&173.9	&8	&12	&0.5000 \\
24	&2008	&4.200	&6	&2014	&113.3	&5	&11	&1.200 \\
25	&2019	&3.600	&——	&——	&——	&——	&——	&—— \\
  \noalign{\smallskip}\hline
\end{tabular*}
\ec
\end{table}

The details of our prediction are as follows: 

About solar cycle 25:
\begin{enumerate}[1)]
\item $R_{max}$, $R_{max}$ of solar cycle 6 is 76.30, and $R_{max}$ of solar cycle 15 is 173.6. Based on the principle of similarity, we can speculate that $R_{max}$ of solar cycle 25 should be between 76.3 and 173.6. In addition, since the corresponding value of solar cycle 25 in the phase of the century cycle is higher than that of solar cycle 24 in the trough, $R_{max}$ of solar cycle 24 is 113.3, and there is an overall enhancement trend over the centuries, we finally predict that $R_{max}$ of solar cycle 25 will be between 113.3 and 180.1, that is to say: $146.7\pm 33.40$;
\item $L$, solar cycle 6 has a period of 13 years and solar cycle 15 has a period of 10 years. In previous analyses of the overall characterization of solar cycle parameters, long cycles of 13 years accounted for only 6.9\% of all cycles, for solar cycle 6 and solar cycle 9, respectively. Solar cycle 6, which began in 1810, is in the historically famous Dalton Minimum (1790 AD - 1830 AD). Based on Table \ref{tab1} and the century cycle phase, the intensity of solar cycle 25 is not predicted to be lower than that of solar cycle 24. Therefore, we tentatively believe that solar cycle 25 will not enter a new minimum similar to the Maunder minimum and the Dalton minimum. Regarding the relationship between the period of solar cycle and the grand minimum, it was pointed out in a study \citep{karak2013} that the period may be prolonged in 1 - 2 solar cycle before entering the grand minimum. Combining these analyses, we believe that the probability of a 13 - year length for solar cycle 25 is very low, and therefore predict that its period $L$ will be about $11\pm 1$ years;
\item $R_{at}$, $R_{at}$ of solar cycle 6 and $R_{at}$ of solar cycle 15 are both less than 1, so we predict that $R_{at}$ of solar cycle 25 is also less than 1, i.e., the rise time is less than the fall time; 
\item $T_{up}$, with the predictions of the period and asymmetric parameter, the rise time predicted in this paper is 4 - 5 year. Since the actual start of solar cycle 25 is 2019 December, we predict that $R_{max}$ of solar cycle 25 will occur in 2024.
\end{enumerate}

About solar cycle 26:
\begin{enumerate}[1)]
\item $R_{max}$ of solar cycle 7 is 117.4, $R_{max}$ of solar cycle 16 is 129.7. we apply the averaged value of solar cycle 7 and 16 to predict solar cycle 26. Considering the increasing trend of solar cycles (the last term of Eq. \eqref{eq2}), we finally predict that $R_{max}$ of solar cycle 26 should be $133.0\pm 3.200$;
\item $L$, the period of both solar cycle 7 and solar cycle 16 is 10 years, so the period of solar cycle 26 is also predicted to be 10 years;
\item $R_{at}$, $R_{at}$ of solar cycle 7 is greater than 1 and $R_{at}$ of solar cycle 16 is equal to 1. Therefore, $R_{at}$ of solar cycle 26 is predicted to be greater than or equal to 1, i.e., the rise time will not be shorter than the fall time;
\item $T_{up}$, the rise time of solar cycle 7 is 7 years and the rise time of solar cycle 16 is 5 years, because $R_{at} (=2.333)$ of solar cycle 7 is the highest of all solar cycles. Considering the period $L$ and asymmetric parameter $R_{at}$, this paper predicts a rise time of 5 - 6 year for solar cycle 26. In the previous section, our prediction for the period $L$ of solar cycle 25 is $11\pm 1$ year, and in predicting solar cycle 26, we take the average of the predicted values for $L$ of solar cycle 25. In other words, we consider the end of solar cycle 25 to be 2030, and the beginning of solar cycle 26 to be 2030. Therefore, we predict that $R_{max}$ of solar cycle 26 will occur between 2035 and 2036;
\item $R_{min}$, according to the fitted empirical function, there is an overall tendency for solar activity to increase, so we predict that $R_{min}$ of solar cycle 26 will be greater than or equal to 3.6 of solar cycle 25.
\end{enumerate}
 
We plotted the prediction curves for solar cycle 25 and 26 as shown in the blue dotted curve in Figure \ref{fig5}. The values are obtained by averaging the previous two solar cycles plus the appropriate solar activity trend. It should be noted that our predictions above do not take into account the influence of the 56.02 - year period. Although the 56.02 - year period is the third strongest on Figure \ref{fig1a}, its presence is not stable (Fig.\ref{fig1b}), and we have reason to suspect that solar cycle 25 is not affected by this period. However, if we consider the influence of the 56.02 - year period, we can also see from Figure \ref{fig5} that the first three century cycles do all fluctuate at their peaks. The troughs are located in solar cycle A and 1 in the century cycle G1, solar cycle 10 in the century cycle G2, and solar cycle 20 in the century cycle G3. However, solar cycle 25 and 26 are located in the ascending phase of the century cycle, so this does not affect our predictions.

The good fit of the actual ASN sequence to the fitted century cycle curve in Figure \ref{fig5} also further validates the correctness of the timescale of the century cycle and the century cycle G3 ending moments that we identified.

\subsubsection{Comparison with Other Predictions}
In addition to the prediction of solar cycle 25, this paper also provides a new prediction for solar cycle 26, which we will compare with the results of the existing predictions of solar cycle 25 and 26 respectively in the followings.

Up to now, many results of solar cycle 25 predictions have been published.\citet{nandy2021}has done a detailed compilation of most of the solar cycle 25 predictions prior to the start of solar cycle 25, since the predictions of solar cycle 25 in this paper are already in the ascending phase of solar cycle 25, we have chosen more predictions made after the beginning of solar cycle 25 (after 2019). We summarize the results of the recent predictions, sorted by the chronological order of the predictions in the different categories of methods (as shown in Table \ref{tab2}). It should be noted in Table \ref{tab2} that there are also differences in the accuracy of the results obtained by different prediction methods, which is also a manifestation of the characteristics of different methods, so we do not make a uniform requirement for accuracy here. Secondly, according to the classification of prediction methods in the previous section, our prediction method belongs to the extrapolation method, which it is listed separately in the table.

Comparing the $R_{max}$ of all predictions in Table \ref{tab2}, we can see that:
\begin{enumerate}[(a)]
\item \citep{guo2021} in the dynamo model category, \citep{diego2021}, \citep{bisoi2020} in the precursor method, \citep{mcintosh2020}, \citep{Sarp2018}in the extrapolation method, and \citep{su2023}, \citep{prasad2022} in the interdisciplinary category are consistent with our results that solar cycle 25 will have a larger $R_{max}$ than solar cycle 24. Among these results, \citep{diego2021} which utilizes the relationship between the recurrence index of geomagnetic activity in the fall phase of the previous cycle and sunspot number in the next cycle for prediction, reaches the same conclusion as ours that the century cycle ends at solar cycle 24;
\item \citep{nagovitsyn2023}, \citep{lu2022}, \citep{xiong2021}, \citep{kumar2021} in the precursor method, and from their predicting results for $R_{max}$, we also believe that their conclusion is that solar cycle 25 is stronger than solar cycle 24;
\item In addition, as in \citep{javaraiah2023} of the precursor method, \citep{javaraiah2023} of the extrapolation method, and the result (20) which is published by the NOAA/NASA co-chaired international panel to forecast Solar Cycle 25 (\url{https://www.swpc.noaa.gov/news/solar-cycle-25-forecast-update}), these predictions of $R_{max}$ are similar to $R_{max}$ of solar cycle 24;
\item \citep{labonville2019}in the dynamo model category, \citep{brajsa2022}, \citep{burud2021}, \citep{chowdhury2021} in the precursor method, \citep{kakad2021} in the extrapolation method, \citep{courtillot2021} in the other category, and \citep{bizzarri2022}, \citep{attia2020} in the interdisciplinary category, lead directly to the conclusion that solar cycle 25 will have a smaller $R_{max}$ than solar cycle 24, or that cycle 25 lies in the trough between century cycles, or that solar cycle 25 will enter a new grand minimum. This is inconsistent with our prediction using the century cycle.
\end{enumerate}

\begin{table}[ht]
\bc
\begin{minipage}[]{100mm}
\caption[]{Comparison of different predictions for the Solar Cycle 25\label{tab2}}
\end{minipage}
\setlength{\tabcolsep}{1pt}
\small
 \begin{tabular*}{\textwidth}{@{\extracolsep{\fill}}c c c c c c @{}}
  \hline\noalign{\smallskip}
  \multicolumn{4}{c}{Forecast Information} & \multicolumn{2}{c}{The Prediction Result of Solar Cycle 25}\\
  \hline \noalign{\smallskip}
  Category& \makecell[c]{Serial\\Number} & \makecell[c]{Time of\\Publication} & Author& $R_{max}$& $t_{max}$ \\
  \hline\noalign{\smallskip}
  \makecell[c]{The Century \\ Cycle}&(1)&2023&This paper& $146.7\pm 33.38$ & 2024 \\
  \hline\noalign{\smallskip}
   \multirow{2}*{\makecell[c]{Dynamo Model \\ Category}} & (2) & 2021 & Guo et al. & 126 & --  \\
		~ & (3) & 2019 & Labonville et al & $89^{+24}_{-14}$ & $2025.3^{+0.89}_{-1.05}$  \\
  \hline\noalign{\smallskip}
   \multirow{11}*{\makecell[c]{Precursor \\ Method}} & (4) & 2023 & Nagovitsyn\& Ivanov & $149\pm 28$ & 2023.5 - 2024.5  \\
		~ & (5) & 2023 & Javaraiah (polar field) & $125\pm 7$ &-- \\
		~ & (6) & 2022 & Brajsa et al. & $121\pm 33$ & -- \\
		~ & (7) & 2022 & Lu et al. & 145.3 & 2024 Oct  \\
		~ & (8) & 2021 & Burud et al. (spotless days)  & $99.13\pm 14.97$ & 2024 Feb - 2024 Mar  \\
		~ & (9) & 2021 & Burud et al. (aa index)  & $104.23\pm 17.35$ & 2024 Feb - 2024 Mar  \\
		~ & (10) & 2021 & Diego\& Laurenza & $205\pm 29$ & around mid-2023  \\
		~ & (11) & 2021 & Xiong et al. & 140.2 & 2024 Mar  \\
		~ & (12) & 2021 & Chowdhury et al. & $100.21\pm15.06$ & 2025 Apr $\pm$ 6.5 months \\
		~ & (13) & 2021 & Kumar et al. & $126\pm 3$ & --  \\
		~ & (14) & 2020 & Bisoi & $133\pm 11$ & -- \\
  \hline\noalign{\smallskip}
   \multirow{4}*{\makecell[c]{Extrapolation \\ Method}} & (15) & 2023 & Javaraiah (sunspot group area) & $125\pm 11$ & --  \\                               
		~ & (16) & 2021 & Kakad\& Kakad & $103.3\pm 15$ & --  \\
		~ & (17) & 2020 & McIntosh et al. & $229\pm 76$ & --  \\
		~ & (18) & 2018 & Sarp et al. & $154\pm 12$ & $2023.2\pm 1.1$ \\				
  \hline\noalign{\smallskip}
   \multirow{2}*{Other Category} & (19) & 2021 & Courtillot et al. & $97.6\pm 7.8$ & $2026.2\pm 1$  \\                               
		~ & (20) & 2019 & NOAA/NASA & $115\pm 10$ & 2025 Jul $\pm$ 8 months  \\
  \hline\noalign{\smallskip}
   \multirow{6}*{\makecell[c]{Interdisciplinary \\ Category}} & (21) & 2023 & Su et al. & $133.9\pm 7.2$ & 2024 Feb    \\                               
		~ & (22) & 2022 & Bizzarri et al. (the TT model)  & $104\pm 7$ & 2024 Jul $\pm$7 months  \\
		~ & (23) & 2022 & Bizzarri et al. (the CV model)  & $110\pm 9$ & 2024 Apr $\pm$12 months  \\
        ~ & (24) & 2022 & Prasad et al. & $171.9\pm 3.4$ & 2023 Aug $\pm$ 2 months \\
        ~ & (25) & 2020 & Attia et al. & $80\pm 12$ & 2026 \\
  \noalign{\smallskip}\hline
\end{tabular*}
\ec
\end{table}

A comparison with the prediction results for cycle 26 is shown in Table \ref{tab3}.

By comparison, we find that the prediction results of $R_{max}$ of solar cycle 26 are quite different in different studies. \citep{kalkan2023} using non-linear autoregressive exogenous neural networks is considered to be similar to or weaker than solar cycle 24. Solar cycle 25 and 26 are believed to be at the minimum of the century cycle, and solar activity will enter a new grand minimum. \citep{liu2023a} using the long short-term memory method shows that solar cycle 25 and 26 are similar, which is different from our result that solar cycle 26 is slightly smaller than solar cycle 25. \citep{becheker2023} using the function fitting method considers solar cycle 26 to be numerically weaker than solar cycle 24. Current projections generally agree that solar cycle 26 will peak between 2035 and 2036.

\begin{table}[ht]
\bc
\begin{minipage}[]{100mm}
\caption[]{Comparison of different predictions for the Solar Cycle 26\label{tab3}}
\end{minipage}
\setlength{\tabcolsep}{1pt}
\small
 \begin{tabular*}{\textwidth}{@{\extracolsep{\fill}}c c c c c c @{}}
  \hline\noalign{\smallskip}
  \multicolumn{3}{c}{Forecast Information} & \multicolumn{2}{c}{The Prediction Result of Solar Cycle 26}\\
  \hline \noalign{\smallskip}
   Serial Number & Time of Publication & Author& $R_{max}$& $t_{max}$ \\
  \hline\noalign{\smallskip}
  (1) & 2023 & This paper& $133.0\pm 3.200$ & 2035 - 2036 \\
  (26) & 2023 & Kalkan et al. & 113.25     &  2036 Oct  \\
  (27) & 2023 & Liu et al. & 135        &  2035 Jan  \\
  (28) & 2023 & Becheker et al. & $96\pm 28$ & --\\

  \noalign{\smallskip}\hline
\end{tabular*}
\ec
\end{table}

\subsection{Discussion}

The ASN sequence is the longest time series data that humans have continuously observed the same celestial body to date, but compared with the history of solar activities its data are still too few and too short. Therefore, it is practically impossible to tell whether the quasi-periodic properties of the solar cycle and the century cycle are due to the incomplete observed cycles showing quasi-periodic properties because of the short time being observed, or whether they are the properties of the solar cycle and the century cycle themselves. To solve this problem, it is necessary to study the forward extension of the sunspot number sequence. The impact of this problem in this paper is that in terms of prediction error, our predictions depend on the peak of the solar cycle at a similar phase of the century cycle to the one being predicted, but there are only two similar-phase solar cycles available for prediction, and so this shows a more different range of error in the predictions for soalr cycle 25 and 26. Our method analyzes long-term trends (centennial scales) in solar activity and is therefore dependent on the length of the data. This is in contrast to the vast majority of prediction methods, such as the precursor method used polar magnetic field, which utilizes only the polar magnetic field of the previous cycle on a scale of about a decade.

The error in the prediction results obtained by different prediction methods varies considerably. For example, the method\citep{mcintosh2020} which has the dependence of prediction results on terminators gives a large error range when undetermined terminators. While the prediction by using comprehensive precursors and multiple regression techniques can significantly improve the prediction accuracy, adaptability and stability, and the regression coefficient can reach 0.95 \citep{xiong2021,lu2022}. As for the prediction results of interdisciplinary (specifically referring to machine learning) category, the goodness of the results lies in the selection and construction of the model, so the error ranges of the prediction results are generally small after the model is determined. However, the quasi-periodicity and complexity of the solar activity make the predictions challenging, and too precise predictions might fall into the risk of overfitting.

In fact, our prediction also includes the effect of the 56.20-year period component in the $R_{max}$ of the solar cycles. This is because we assume that cycles located in similar phases on the background of the century cycle will have similar parameter characteristics, while the 56.20-year period component is only implied in the relatively small Schwabe cycle of the peak year of century cycle. Figure \ref{fig5} shows that cycle A in G1, cycle 10 in G2 and cycle 20 in G3 are located near the peak year of the century cycles, but they are similarly relatively weak. Based on their statistical trend, we may also predict the future cycle located near the peak year of G4. Obviously, this component has little impact on the prediction of the solar cycle 25 and 26 which are located in the early ascending phase of the century cycle.

With regard to the comparison of results, according to 3.1.4 we can easily realize that the published predictions are all very significantly different from each other, presenting a chaotic state. We believe that this is a clear manifestation of the complex and quasi-periodic nature of the solar cycle. Although there is a huge difference between the results, a comparison of the results obtained from the different methods of prediction can also give us a deeper understanding of the solar cycle. We also find that of all the predictions, the later the prediction, the greater the probability of the result of the work that believes that solar cycle 25 will be stronger than solar cycle 24, which is also consistent with our predictions. One of the reasons for this is that the later the prediction is made, the more pronounced the evolutionary trend exhibited by the relative sunspot number in solar cycle 25, but it also illustrates that none of the current types of prediction methods have a clear advantage in predicting long time in advance, with the most constrained being the precursor method. Unlike these methods, our prediction method is based on the idea that solar cycles in similar phases of the century cycle have similar cycle parameters. Therefore, after determining the existence of the century cycle, as well as the shape (fitting empirical function) and timescale of the century cycles (parameter $C$), our prediction is not basically limited by the early or late prediction time. The closer the time of our prediction is to the time at $t_{max}$, the closer our prediction will be to the true values, without affecting our prediction of the long-term trend of the solar cycle. Our method has a requirement for data length, but there are two sides to everything. On earlier prediction, our work is superior.

Among the many results of predictions for solar cycle 25, many of those that thought solar cycle 25 would be smaller than solar cycle 24, contrary to our prediction, and they thought that there would be a new grand minimum in solar cycle 25 similar to the Maunder minimum and the Dalton minimum. However, looking at the current data for solar cycle 25, the probability of it being weaker than solar cycle 24 is dramatically reduced. In other words, the trough between century cycles G3 and G4 does not have a very pronounced long-lasting grand minimum, and we can probably assume that the correlation between century cycles and grand minimum is weakened. But how exactly this will actually play out, and whether a new grand minimum will emerge, still depends on how the sunspot number data develop.

\section{Conclusion}

This work confirms the existence of solar century cycles, determines their main parametric characteristics, and obtains a empirical function (Eq.(\ref{eq2})).  

\begin{enumerate}[(1)]
\item The century cycles do exist and their period $L$ is 104.0 years;
\item Solar activity has a gradually enhancing trend on the timescale of century cycles.
\item The solar Schwabe cycle 24 is located in the valley between the century cycles G3 and G4. After that, cycle 25 and 26 are in the ascending phase of the century cycle G4, therefore, a new grand minimum will not occur. Or, that is to say, the grand minimum just occurred around cycle 24.  
\end{enumerate}

The existence of the solar century cycle indicates that there must be a global physical mechanism on the Sun that is larger in scale than the solar Schwabe cycle, influencing the patterns of solar activity. But so far, we are not yet understanding this global physical mechanism. The century cycle pattern can help us predict the basic characteristics of forthcoming Schwabe cycles. Here, based on the empirical functions of solar century cycle, and the assumption of similarity extrapolation, we predict the main parametric characteristics of the solar cycles 25 and 26. The predicting results are as follows:

\begin{enumerate}[(1)]
\item Solar cycle 25 will reach its maximum in 2024, with a peak ASN $146.7\pm 33.40$, stronger than cycle 24, and its period is about $11\pm 1$ years;
\item Solar cycle 26 will start from 2030, reach to its maximum between 2035 and 2036 with a peak ASN $133.0\pm 3.200$, and its period is about 10 years.
\end{enumerate}

This paper ignores some sunspot anomaly data when examining the century cycle of the solar activity. This may result in our predictions of the long - term trend not being wrong, but there are some uncertainties in the details. It is possible that future studies will be able to interpret the sunspot anomaly data and combine multiple solar cycle features for future cycle predictions to improve accuracy. In addition, in studying the long-term evolutionary characteristics of solar activity, long-term data on isotopes (especially carbon - 14 isotopes) on Earth could be analyzed in addition to observations of the relative sunspot number. This has the potential to extend the time series to thousands of years in length and give a more accurate picture of the solar activity long-term evolutionary pattern.

\begin{acknowledgements}
This work is supported by the National Key R\&D Program of China 2021YFA1600503, 2022YFF0503001, the Strategic Priority Research Program of the Chinese Academy of Sciences XDB0560302, NSFC Grant 11973057, and the International Partnership Program of Chinese Academy of Sciences 183311KYSB20200003.
\end{acknowledgements}

\bibliographystyle{raa}
\bibliography{msRAA-2023-0354.R1}

\end{document}